\documentstyle[12pt,aaspp4,flushrt,eqsecnum,epsfig]{article}

\newcommand{\ev}{\mbox{\rm \,eV}}

\newcommand{\kev}{\mbox{\rm \,keV}}

\newcommand{\dtri}{d^{\,3}}
\newcommand{\thte}{\theta_{\!e}}
\newcommand{\thtr}{\theta_{\!r}}
\newcommand{\ord}{\mbox{O}}
\newcommand{\simlt}{\lesssim}
\newcommand{\simgt}{\gtrsim}
\newcommand{\eqref}[1]{(\ref{#1})} 
\newcommand{\tfrac}[2]{\case{#1}{#2}}
\newcommand{\half}{\case{1}{2}}
\newcommand{\et}[1]{e^{\mbox{\footnotesize $#1$}}}

\begin{document}

\title{RELATIVISTIC CORRECTIONS TO THE SUNYAEV-ZEL'DOVICH EFFECT}
\author{Anthony Challinor\footnote{A.D.Challinor@mrao.cam.ac.uk} \&\ Anthony Lasenby\footnote{A.N.Lasenby@mrao.cam.ac.uk}}
\affil{MRAO, Cavendish Laboratory, Madingley Road, Cambridge, CB3 0HE, U.K.}

\begin{abstract}

We present an extension of the Kompaneets equation which allows relativistic
effects to be included to any desired order. Using this, we are able to obtain
simple analytic forms for the spectral changes due to the Sunyaev-Zel'dovich
effect in hot clusters, correct to first and second order in the expansion
parameter $\thte=k_{B} T_{e}/m c^2$. These analytic forms agree with previous
numerical calculations of the effect based upon the multiple scattering
formalism, and are expected to be very accurate over all regions of the CMB
spectrum for $k_{B} T_e$ up to $\sim 10 \kev$. Our results confirm previous
conclusions that the result of including relativistic corrections in the
Sunyaev-Zel'dovich effect is a small {\em reduction\/} in the amplitude of the
effect over the majority of the spectrum: specifically we find $\Delta T/T = -2
y ( 1- 17/10 \,\thte +123/40 \,\thte^2)$ (correct to second-order) in the
Rayleigh-Jeans region, where $y$ is the usual Comptonization parameter. For a
typical cluster temperature of $8 \kev$, this amounts to a correction downwards
to the value of the Hubble constant derived using combined X-ray and
Rayleigh-Jeans Sunyaev-Zel'dovich information by about 5 percent.

\end{abstract}

\keywords{cosmology: cosmic microwave background --- galaxies: clusters: general --- radiative transfer --- scattering}

\section{Introduction}

Non-relativistic treatments of the Sunyaev-Zel'dovich effect usually employ
the Kompaneets equation~\cite{komp57} to determine the distortion of the
Cosmic Microwave Background (CMB) spectrum.
The Kompaneets equation does not, however, include relativistic effects, which
may be important for hot clusters where $k_{B}T_{e} \simgt 10 \kev$.
For this reason, and because of the low optical depth of typical clusters,
relativistic treatments of the Sunyaev-Zel'dovich effect usually employ
a multiple scattering description of the
Comptonization process~\cite{wright79,fabbri81,taylor89,loeb91,reph95}.
Including
relativistic effects in this procedure gives a complicated expression
for the spectral distortion, which is best handled by numerical
techniques (see, for example,~\cite{reph95}).

In this paper, we show how the Kompaneets equation may be extended to
include relativistic effects in a self-consistent manner,
allowing the Sunyaev-Zel'dovich effect in
hot clusters to be described on the basis of a Kompaneets type equation.
The extension of the Kompaneets equation can be carried out to
arbitrary orders in relativistic effects,
although we shall only consider the lowest order corrections here. The
resulting equation conserves photons at every order, and we demonstrate
that it is consistent with earlier calculations of the energy transfer rate
between the plasma and a Planck distribution of photons~\cite{woodward70}.

We then consider the application of the generalised Kompaneets equation
to the calculation of the Sunyaev-Zel'dovich effect in hot clusters.
Simple analytic forms are given for the spectral distortions in the
limit of small optical depth, including relativistic effects to
second-order. These are
in excellent agreement with Rephaeli's (1995) numerical calculations,
which were based on the multiple scattering approach (truncated at one
scattering). This lends further support to Fabbri's observation
that the Boltzmann equation can be applied to describe the Sunyaev-Zel'dovich
effect in optically thin clusters~\cite{fabbri81}, despite claims to the
contrary~\cite{wright79}.

The relativistic corrections to the Sunyaev-Zel'dovich effect are of
importance in the calculation of the Hubble constant $H_{0}$ by
the Sunyaev-Zel'dovich route in hot clusters (see, for
example,~\cite{las97} for a recent review of the Sunyaev-Zel'dovich route
of determining $H_{0}$, and~\cite{saun96} for recent observations).
In the Rayleigh-Jeans region, we find that
relativistic effects lead to a small decrease in the Sunyaev-Zel'dovich effect,
and hence a small reduction in the hitherto determined values of $H_{0}$, in
agreement with the conclusions in~\cite{reph97}.

We employ natural units, $c=\hbar=1$, except in the discussion of the
Sunyaev-Zel'dovich effect in Section~\ref{sz_sec_sun_zel}.

\section{Extending the Kompaneets Equation}
\label{sz_sec_komp}

In this paper we shall not consider effects due to the peculiar motion of the
cluster (such effects give rise to a kinetic correction to the
Sunyaev-Zel'dovich effect). For a comoving cluster, the CMB photon distribution
function is isotropic and may be denoted $n(\omega)$, where $\omega$ is the
photon energy. The electrons are assumed to be in thermal equilibrium at
temperature $T_{e}$, and are described by an isotropic distribution
function $f(E)$, where $E$ is the photon energy.
The Boltzmann equation
describing the evolution of $n(\omega)$ may be written as~\cite{buchler76}
\begin{equation}
\frac{\partial n(\omega)}{\partial t} = -2 \int \frac{\dtri p}{(2\pi)^{3}}
\dtri p' \dtri k' \,W \Bigl[n(\omega)\bigl(1+ n(\omega')\bigr)f(E) -
n(\omega')\left(1+n(\omega)\right)f(E') \Bigr],
\label{sz_eq_1}
\end{equation}
where $W$ is the invariant transition amplitude for Compton
scattering of a photon of 4-momentum $k^{\mu}$
by an electron (of charge $e$ and mass $m$) with
4-momentum $p^{\mu}$, to a photon momentum $k^{\prime\mu}$ and an
electron momentum $p^{\prime\mu}$~\cite{ber-quan}:
\begin{eqnarray}
W &=& \frac{(e^{2}/4\pi)^{2} \bar{X}}{2\omega\omega' E E'}
\delta^{4}(p+k-p'-k') \label{sz_eq_2} \\
\bar{X} &\equiv& 4m^{4} \left(\frac{1}{\kappa} + \frac{1}{\kappa'}\right)^{2} -
4m^{2} \left(\frac{1}{\kappa} + \frac{1}{\kappa'}\right) -
\left(\frac{\kappa}{\kappa'} + \frac{\kappa'}{\kappa}\right),\label{sz_eq_2b}
\end{eqnarray}
with $\kappa \equiv -2p^{\mu}k_{\mu}$ and $\kappa'\equiv
2p^{\mu}k^{\prime}_{\mu}$. In equation~\eqref{sz_eq_1},
we have assumed that electron
degeneracy effects may be ignored.

The electrons are described by a relativistic Fermi distribution. Since we
are ignoring degeneracy effects, we have
\begin{equation}
f(E)\approx \et{-(E-\mu)/k_{B}T_{e}}.
\end{equation}
Substituting this form for $f(E)$ into equation~\eqref{sz_eq_1},
and expanding the
term in brackets in the integrand in powers of $\Delta x$, where
\begin{eqnarray}
x & \equiv & \frac{\omega}{k_{B}T_{e}} \\
\Delta x &\equiv & \frac{\omega'-\omega}{k_{B}T_{e}},
\end{eqnarray}
gives a Fokker-Planck expansion
\begin{eqnarray}
\lefteqn{\frac{\partial n(x)}{\partial t} = 2
\left(\frac{\partial n}{\partial x} + n(1+n)\right) I_{1} 
+ 2 \left(\frac{\partial^{2}n}{\partial x^{2}} +
2(1+n)\frac{\partial n}{\partial x} + n(1+n) \right) I_{2} }
\nonumber \\
& & \hspace{1.6cm} \mbox{} + 2\left(\frac{\partial^{3}n}{\partial x^{3}}+3(1+n)
\frac{\partial^{2}n}{\partial x^{2}} + 3 (1+n) 
\frac{\partial n}{\partial x} + n(1+n) \right) I_{3}
\nonumber \\
& & \hspace{-1.5cm} \mbox{} + 2\left(\frac{\partial^{4} n}
{\partial x^{4}} + 4(1+n)
\frac{\partial^{3} n}{\partial x^{3}} + 6(1+n)
\frac{\partial^{2} n}{\partial x^{2}} + 4(1+n)
\frac{\partial n}{\partial x} + n(1+n) \right) I_{4} + \cdots, 
\label{sz_eq_3}
\end{eqnarray}
where
\begin{equation}
I_{n} \equiv
\frac{1}{n!}\int \frac{\dtri p}{(2\pi)^{3}} \dtri p' \dtri k' \,W f(E)
(\Delta x)^{n},
\label{sz_eq_4}
\end{equation}
which does not depend on $n(\omega)$.

The calculation of the $I_{n}$ may be performed by
expanding the integrand in powers of $p/m$ and $\omega/m$. The factor
$f(E)$ is handled by the expansion
\begin{equation}
f(E) \approx \et{(\mu-m)/k_{B}T_{e}} \et{-u}
\Bigl(1+\half \thte u^{2} + \tfrac{1}{8}\thte^{2}u^{3}(u-4) + \cdots\Bigr),
\label{sz_eq_8}
\end{equation}
where
\begin{eqnarray}
u &\equiv& \frac{p^{2}}{2mk_{B}T_{e}} \\
\thte &\equiv& \frac{k_{B}T_{e}}{m},
\label{sz_eq_9}
\end{eqnarray}
and the chemical potential $\mu$ may be eliminated by introducing the
electron number density $N_{e}$, which evaluates to
\begin{equation}
N_{e} = \frac{m\sqrt{2m}}{\pi^{2}} \Gamma(3/2)
(k_{B}T_{e})^{\mbox{\scriptsize $\frac{3}{2}$}} \et{(\mu-m)/k_{B}T_{e}}
\left(1 + \frac{15}{8} \thte + \frac{105}{128} \thte^{2} + \ord(\thte^{3})
\right).
\label{sz_eq_10}
\end{equation}
Note that equation~\eqref{sz_eq_10} is an asymptotic expansion about $\thte=0$
of the result
\begin{equation}
N_{e} = \frac{m^{3}}{\pi^{2}} \thte K_{2} (1/\thte) \et{\mu/k_{B}T_{e}},
\end{equation}
where $K_{2}(x)$ is a modified Bessel function, which suggests that our
series expansions of the $I_{n}$ will only be asymptotic series.
The calculations of the $I_{n}$ are ideally suited to symbolic computer
algebra packages (we use Maple).
Expressing the results in terms of $\thte$, $x$, and the Thomson
cross section $\sigma_{T}$, we find
\begin{eqnarray}
I_{1} & = & \half \sigma_{T} N_{e} \thte x \left( (4-x) +
\thte(10 - \tfrac{47}{2}
x + \tfrac{21}{5}x^{2} ) \right) + \ord(\thte^{3})
\nonumber \\
I_{2} & = & \half \sigma_{T} N_{e} \thte x^{2} \left(1 + \thte(\tfrac{47}{2}
-\tfrac{63}{5} x + \tfrac{7}{10} x^{2} ) \right) + \ord(\thte^{3})
\nonumber \\
I_{3} & = & \half \sigma_{T} N_{e} \thte x^{3} \left(\tfrac{7}{5}(6-x) \thte
\right) + \ord(\thte^{3}) \nonumber \\
I_{4} & = &\half \sigma_{T} N_{e} \thte x^{4} (7\thte ) + \ord(\thte^{3}).
\label{sz_eq_is}
\end{eqnarray}
For $n>4$, $I_{n}$ is third-order or higher in $\thte$.
For CMB photons passing through a cluster at
redshift $z$, we have $\bar{x}\simeq 6.2\times 10^{-4}(1+z)/k_{B}T_{e}$,
where $\bar{x}$ is the average of $x$ over a Planck distribution and
$k_{B}T_{e}$ expressed in $\ev$. The electron temperature is typically
$\simlt 10\kev$, so that $x\ll 1$.

Substituting these expansions for the $I_{n}$ into the
series~\eqref{sz_eq_3}, we develop an expansion of $\partial n/\partial t$
in $\thte$. In this paper, we shall only be
concerned with the lowest order relativistic corrections, and start by
retaining all terms up to $\ord(\thte^{2})$. To include all such terms
consistently, it is necessary to retain only the first four terms in the
series~\eqref{sz_eq_3}. A lengthy calculation gives the result
\begin{equation}
\frac{\partial n(x)}{\partial t} = \sigma_{T} N_{e}\thte
\frac{1}{x^{2}}\frac{\partial}
{\partial x} \left(x^{2} j(x)\right),
\label{sz_eq_phcons}
\end{equation}
where the current $j(x)$ is given by
\begin{eqnarray}
j(x) & = & x^2\Biggl[\left(\frac{\partial n}{\partial x} +
n(1+n)\right)
+ \thte \Biggl[ \frac{5}{2} \left(\frac{\partial n}{\partial x} + n(1+n)
\right) + \frac{21}{5} x \frac{\partial}{\partial x}\left(\frac{\partial n}
{\partial x} + n(1+n)\right)  \nonumber \\
& & \mbox{} + \frac{7}{10} x^2 
\left( \frac{\partial^{3} n}{\partial x^{3}} +
2\frac{\partial^{2} n}{\partial x^{2}}(1+2n) + \frac{\partial n}{\partial x}
\left(1-2\frac{\partial n}{\partial x}\right)
\right) \Biggr]+ \ord(\thte^{2}) \Biggr].
\label{sz_eq_cur}
\end{eqnarray}
The zero-order term in equation~\eqref{sz_eq_cur} is just that term
which usually
appears in the Kompaneets equation~\cite{komp57}. The $\ord(\thte)$ term is
the lowest-order relativistic correction to the current. The form
of equation~\eqref{sz_eq_phcons} ensures conservation of the total
number of photons, which is true for each order in $\thte$. A similar
expression has been derived independently by Stebbins (1997) using
non-covariant methods, although he only considers the lowest-order terms
in $\omega/m$.

This derivation makes it clear that trying to include relativistic terms
in the framework of the usual Kompaneets equation~\cite{komp57},
although it would seem natural to retain only terms in $I_{1}$ and $I_{2}$,
would be incorrect, since it would
not obey photon conservation. To be specific, if only $I_{1}$ and $I_{2}$ are
retained, then photon conservation requires that the current be given by
\begin{equation}
j(x) = 2 \frac{I_{2}}{\sigma_{T}N_{e}\thte}
\left(\frac{\partial n}{\partial x} + n(1+n)\right), 
\end{equation}
and consistency with the expansion of the Boltzmann equation~\eqref{sz_eq_3}
requires that
\begin{equation}
I_{1} = \frac{\partial I_{2}}{\partial x} + 2\frac{I_{2}}{x} - I_{2}.
\label{sz_eq_cons}
\end{equation}
Calculating $I_{2}$ to $\ord(\thte^{2})$ in the limit of
small $x$, gives the current $j'(x)$ (in the limit of small $x$) where
\begin{equation}
j'(x) = x^2 \left(\frac{\partial n}{\partial x} + n(1+n) \right) \left(
1 + \frac{47}{2} \thte \right),
\label{sz_eq_wrong}
\end{equation}
which amounts to a relativistic correction to the cross section appearing
in the usual Kompaneets equation. This result is incorrect, since the
approach is not self-consistent to the order that the answer is quoted.
To see this, one need only calculate $I_{1}$ using equation~\eqref{sz_eq_cons}
and the expression for $I_{2}$ from equation~\eqref{sz_eq_is}. The expression
for $I_{1}$ obtained by this procedure only agrees with the direct
calculation equation~\eqref{sz_eq_is} to $\ord(\thte)$,
not to $\ord(\thte^{2})$.
One obtains a better approximation to the true
current (eq.~\eqref{sz_eq_cur}) by calculating $I_{1}$ directly and then
using equation~\eqref{sz_eq_cons} to calculate an effective value of
$I_{2}$, which may
then be used to deduce the current. This procedure reproduces the first term
in the $\ord(\thte)$ correction in $j(x)$, in the limit of small $x$, and
is similar to that employed by Fabbri~(1981) in obtaining his
equation (11).
However, it is only the full expression equation~\eqref{sz_eq_cur} which is
consistent with the original Boltzmann equation~\eqref{sz_eq_1} to
$\ord(\thte^{2})$.

\section{Rate of Energy Transfer}

As a check on the consistency of equation~\eqref{sz_eq_cur} with
existing results
in the literature, we calculate the energy transfer rate between the
electrons and a Planck distribution of photons at temperature $T_{r}$.

Multiplying the continuity equation~\eqref{sz_eq_phcons} by $x^{3}$ and
integrating, we find
\begin{equation}
\frac{\partial}{\partial t} \int_{0}^{\infty} x^{3} n(x) \, dx =
- \sigma_{T} N_{e}\thte \int_{0}^{\infty} x^{2} j(x) \, dx.
\label{sz_eq_etrans_i}
\end{equation}
The left-hand side of equation~\eqref{sz_eq_etrans_i} is proportional to the
rate at which the
photons are gaining energy per unit volume, denoted $dE/dt$.
Substituting a Planck distribution for $n(\omega)$ in
equation~\eqref{sz_eq_cur} and integrating gives the result
\begin{equation}
\frac{dE_{r}}{dt} = 4E_{r}N_{e} \sigma_{T} (\thte-\thtr)
\left(1+ \frac{5}{2} \thte - 21\frac{\zeta(6)}{\zeta(4)} \thtr +
\ord(\theta^{2}) \right),
\label{sz_eq_etrans}
\end{equation}
where
\begin{equation}
\thtr \equiv \frac{k_{B}T_{r}}{m},
\end{equation}
and $\zeta(x)$ is the Riemann Zeta function.

This result may be compared with a direct evaluation of the energy transfer,
obtained by multiplying the transition rate $W$ by the
energy transfer $\omega'-\omega$, and integrating over all collisions:
\begin{equation}
\frac{dE_{r}}{dt} = 4 \int \frac{\dtri k}{(2\pi)^{3}} \frac{\dtri k'}
{(2\pi)^{3}}
\dtri p\, \dtri p' \, W f(E) n(\omega)[1+ n(\omega')] (\omega'-\omega).
\label{sz_eq_17}
\end{equation}
The integral may be evaluated by a consistent expansion of the integrand.
A lengthy calculation gives the result~\eqref{sz_eq_etrans}, which also
agrees with the non-covariant calculation of Woodward~(1970) (who
gives higher-order corrections also). Note that if we had considered only
$I_{1}$ and $I_{2}$, then using $j'(x)$ given by equation~\eqref{sz_eq_wrong},
we would have obtained (in the limit of small $\thtr$)
\begin{equation}
\frac{dE_{r}}{dt} = 4E_{r}N_{e}\sigma_{T} (\thte-\thtr)\left(
1 + \frac{47}{2}\thte + \ord(\thte^{2}) \right),
\end{equation}
which grossly overestimates the importance of the relativistic corrections.

\section{The Sunyaev-Zel'dovich Effect}
\label{sz_sec_sun_zel}

In this section we apply the generalised Kompaneets equation (to
first-order in relativistic corrections) to the
calculation of the Sunyaev-Zel'dovich effect in optically thin clusters.
We consider higher-order effects in the next section.

Following the standard assumptions, we assume that the optical depth
is sufficiently small that the spectral distortions are small. In this limit,
we may solve equation~\eqref{sz_eq_phcons} iteratively.
The lowest order solution is obtained by substituting the initial photon
distribution $n_{0}(x)$ into the current (eq.~\eqref{sz_eq_cur}).
The integral over time is then trivial, and
may be replaced by an integral along the line of sight through the cluster,
giving
\begin{equation}
\Delta n(x) = \frac{y}{x^{2}} \frac{\partial}{\partial x} \left(x^{2} j(x)
\right),
\label{sz_eq_z1}
\end{equation}
where $j(x)$ is evaluated with $n_{0}(x)$, and
\begin{equation}
y \equiv \sigma_{T} \int N_{e} \thte \, dl,
\label{sz_eq_z2}
\end{equation}
where the integral is taken along the line of sight through the cluster.

For the CMB we take the initial (undistorted) photon distribution to be
Planckian with temperature $T_{0}$:
\begin{equation}
n_{0}(x) = \frac{1}{\et{\alpha x} - 1},
\label{sz_eq_z3}
\end{equation}
where $\alpha \equiv T_{e}/T_{0}$ is the (large) ratio of electron temperature
to the CMB temperature. Evaluating equation~\eqref{sz_eq_z1} in the limit of
large $\alpha$, we find the following fractional distortion:
\begin{eqnarray}
\frac{\Delta n(X)}{n(X)} & = & \frac{yX\et{X}}{\et{X}-1} \Biggl[
X\coth(\half X) - 4 + \thte \Bigl[
-10 + \frac{47}{2} X \coth(\half X) - \frac{42}{5} X^2 \coth^{2}(\half X)
\nonumber
\\
&& \mbox{} + \frac{7}{10} X^3 \coth^{3}(\half X) +
 \frac{7X^{2}}{5\sinh^{2}(\half X)}
\left(X \coth(\half X) - 3\right) \Bigr] \Biggr],
\label{sz_eq_z4}
\end{eqnarray}
correct to first-order in relativistic effects, where
\begin{equation}
X \equiv \frac{\hbar\omega}{k_{B} T_{0}}.
\end{equation}
The first two terms in square brackets in equation~\eqref{sz_eq_z4} give the
usual non-relativistic Sunyaev-Zel'dovich expression, while the terms
proportional to $\thte$ are the lowest order relativistic correction.
Equation~\eqref{sz_eq_z4} agrees with the result in Stebbins (1997).
In the Rayleigh-Jeans limit (small $X$), we find
\begin{equation}
\frac{\Delta n(X)}{n(X)} \simeq -2 y
\left(1- \frac{17}{10} \thte + \ord(\thte^{2})\right).
\end{equation}
%

%%%%%%%%%%%%%%%%%%%%%%%%%%%%%
\begin{figure}[t!]
\begin{center}
\begin{picture}(340,210)
 
%\multiput(0,0)(10,0){35}{\line(0,1){210}}
%\put(0,0){\makebox(0,0)[t]{\tiny 0}}
%\put(50,0){\makebox(0,0)[t]{\tiny 50}}
%\put(100,0){\makebox(0,0)[t]{\tiny 100}}
%\put(0,0){\makebox(0,0)[r]{\tiny 0}}
%\put(0,50){\makebox(0,0)[r]{\tiny 50}}
%\put(0,100){\makebox(0,0)[r]{\tiny 100}}
%\multiput(0,0)(0,10){22}{\line(1,0){340}}
\put(-5,235){\hbox{\epsfig{figure=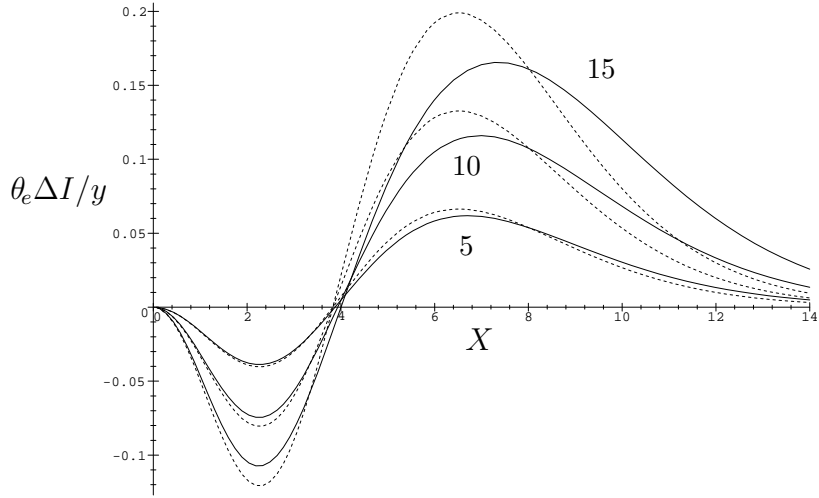,angle=-90,width=12cm}}}
\put(165,73){\makebox(0,0)[t]{\normalsize $X$}}
\put(24,125){\makebox(0,0)[r]{\normalsize $\thte\Delta I/y$}}
\put(160,108){\makebox(0,0)[t]{\small $5$}}
\put(160,138){\makebox(0,0)[t]{\small $10$}}
\put(211,175){\makebox(0,0)[t]{\small $15$}}
\end{picture}
\end{center}
\caption{\sl{The intensity change $\thte\Delta I/y$
(in units of $2(k_{B}T_{0})^{3}
/(hc)^{2}$) plotted against $X$ for three values
of $k_{B}T_{e}$ (in $\kev$).
The solid curves are calculated using the first-order
correction to the Kompaneets equation, while the dashed lines are
calculated from the usual Kompaneets expression.}}
\label{sz_fig_1}
\end{figure}
%%%%%%%%%%%%%%%%%%%%%%%%%%%%%%%%%%%

In Figure~\ref{sz_fig_1} we plot the change in spectral intensity
$\thte\Delta I/y$ as a function of $X$, where
\begin{equation}
\Delta I = \frac{X^{3}}{\et{X}-1} \frac{\Delta n}{n}.
\end{equation}
Also plotted in Figure~\ref{sz_fig_1} are the non-relativistic predictions
made with the standard Kompaneets equation. 
The curves in Figure~\ref{sz_fig_1} are for $k_{B}T_{e} = 5$, $10$, and 
$15\kev$, which are the same as the parameters used
by Rephaeli~(1995) in his Figure 1. His calculations, which
were based on the multiple scattering formalism~\cite{wright79}
and required a numerical analysis, give results in excellent agreement
with ours, which only require the use of the simple expression~\eqref{sz_eq_4}.
This suggests that there is no problem in principle with applying the Boltzmann
equation to the problem of Comptonization in clusters even though the
optical depth may be very small. Similar conclusions were reached by
Fabbri~(1981), but his demonstration was restricted to low
temperature clusters where relativistic effects are not important.

It is clear from Figure~\ref{sz_fig_1} that for $X \simlt 8$, the relativistic
corrections lead to a reduction in the magnitude of the intensity change,
compared to the non-relativistic prediction. This in turn leads to
a reduction in the inferred value of the Hubble constant determined by the
Sunyaev-Zel'dovich route.

%%%%%%%%%%%%%%%%%%%%%%%%%%%%%%%%%%%
 
\begin{figure}[t!]
\begin{center}
\begin{picture}(340,210)
 
%\multiput(0,0)(10,0){35}{\line(0,1){210}}
%\put(0,0){\makebox(0,0)[t]{\tiny 0}}
%\put(50,0){\makebox(0,0)[t]{\tiny 50}}
%\put(100,0){\makebox(0,0)[t]{\tiny 100}}
%\put(0,0){\makebox(0,0)[r]{\tiny 0}}
%\put(0,50){\makebox(0,0)[r]{\tiny 50}}
%\put(0,100){\makebox(0,0)[r]{\tiny 100}}
%\multiput(0,0)(0,10){22}{\line(1,0){340}}
 
\put(-5,235){\hbox{\epsfig{figure=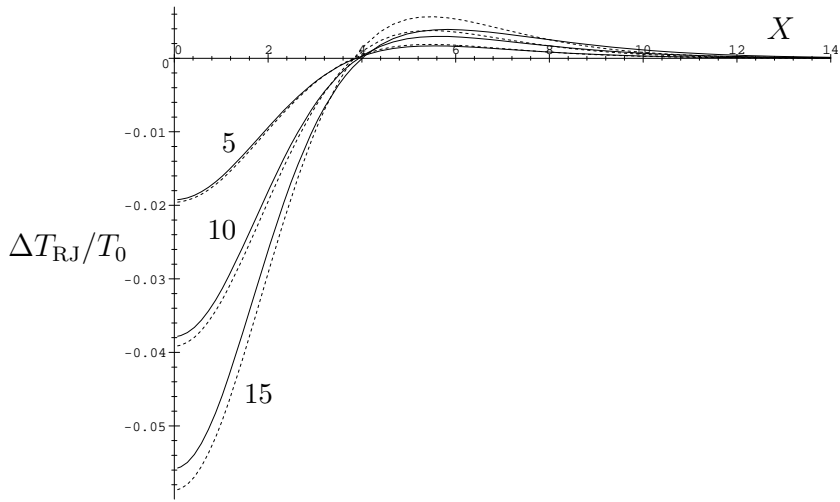,angle=-90,width=12cm}}}
\put(271,184){\makebox(0,0)[b]{\normalsize $X$}}
\put(24,105){\makebox(0,0)[r]{\normalsize $\Delta T_{\mbox{\scriptsize RJ}}/T_{0
}$}}
\put(65,145){\makebox(0,0)[r]{\small $5$}}
\put(65,112){\makebox(0,0)[r]{\small $10$}}
\put(68,50){\makebox(0,0)[l]{\small $15$}}
  
\end{picture}
\end{center}
\caption{\sl{The fractional change $\Delta T_{\mbox{\scriptsize RJ}} / T_{0}$
in the Rayleigh-Jeans brightness temperature plotted against $X$ for three
values of $k_{B}T_{e}$ (in $\kev$). The solid curves are
calculated using the first-order
correction to the Kompaneets equation, while the dashed lines are
calculated from the usual Kompaneets expression.}}
\label{sz_fig_2}
\end{figure}
 
%%%%%%%%%%%%%%%%%%%%%%%%%%%%%%%%%%%%%%%%

In Figure~\ref{sz_fig_2} we plot the fractional change in the Rayleigh-Jeans
brightness temperature $\Delta T_{\mbox{\scriptsize RJ}}/T_{0}$ (divided
by $y/\thte$), where
\begin{equation}
\frac{\Delta T_{\mbox{\scriptsize RJ}}}{T_{0}} = \frac{X}{\et{X}-1}
\frac{\Delta n}{n},
\end{equation}
for the same $\thte$ as in Figure~\ref{sz_fig_1}. The
relativistic corrections to the change in the Rayleigh-Jeans brightness
temperature are significant even at low frequency, unlike the corrections
to the intensity, where relativistic corrections are small in the
Rayleigh-Jeans part of the spectrum.

\section{Higher-order Effects}

We have found that for $k_{B}T_{e} \simgt 10 \kev$ the second-order
relativistic effects make a significant contribution to the
spectral distortion, while third-order effects are only significant for
$k_{B}T_{e} \simgt 15 \kev$.

These calculations require a straightforward extension of the method
of Section~\ref{sz_sec_komp} to include terms at $\ord(\thte^{3})$ (for
second-order relativistic effects). For the calculation to $\ord(\thte^{3})$,
it is necessary to retain the first six
terms of the series~\eqref{sz_eq_3}, and to calculate $I_{1}$ through $I_{6}$
to $\ord(\thte^{3})$. The first iteration
of equation~\eqref{sz_eq_phcons} for $T_{e} \gg T_{0}$
gives the following next order (in $\thte$) correction to $\Delta n/n$:
\begin{eqnarray}
\left(\frac{\Delta n(X)}{n(X)}\right)^{\!(2)} &=&
\thte^{2} \frac{yX\et{X}}{\et{X}-1} \Bigl[-\frac{15}{2} + \frac{1023}{8}
X\coth(\half X) - \frac{868}{5} X^{2} \coth^{2}(\half X) 
\nonumber
\\
&& \mbox{} \hspace{-1.5cm} + \frac{329}{5}
X^{3}\coth^{3}(\half X) - \frac{44}{5} X^{4} \coth^{4}(\half X) +
\frac{11}{30} X^{5} \coth^{5}(\half X)
\nonumber \\
&& \hspace{-3.5cm}\mbox{} + \frac{X^{2}}{30 \sinh^{2}(\half X)}
\left( -2604 + 3948 X\coth(\half X) 
- 1452 X^{2} \coth^{2}(\half X) + 143
X^{3} \coth^{3}(\half X) \right) 
\nonumber\\
&& \mbox{} \hspace{2.3cm} + \frac{X^{4}}{60\sinh^{4}(\half X)}
\left( -528 + 187 X\coth(\half X) \right) \Bigr].
\label{sz_eq_s_1}
\end{eqnarray}
In the Rayleigh-Jeans limit, we find
\begin{equation}
\frac{\Delta n(X)}{n(X)} \simeq -2y \left( 1- \frac{17}{10} \thte +
\frac{123}{40} \thte^{2} + \ord(\thte^{3}) \right).
\label{sz_eq_s_2}
\end{equation}
%

%%%%%%%%%%%%%%%%%%%%%%%%%%%%%%%%%%%
 
\begin{figure}[t!]
\begin{center}
\begin{picture}(340,210)
 
%\multiput(0,0)(10,0){35}{\line(0,1){210}}
%\put(0,0){\makebox(0,0)[t]{\tiny 0}}
%\put(50,0){\makebox(0,0)[t]{\tiny 50}}
%\put(100,0){\makebox(0,0)[t]{\tiny 100}}
%\put(0,0){\makebox(0,0)[r]{\tiny 0}}
%\put(0,50){\makebox(0,0)[r]{\tiny 50}}
%\put(0,100){\makebox(0,0)[r]{\tiny 100}}
%\multiput(0,0)(0,10){22}{\line(1,0){340}}
 
\put(-5,235){\hbox{\epsfig{figure=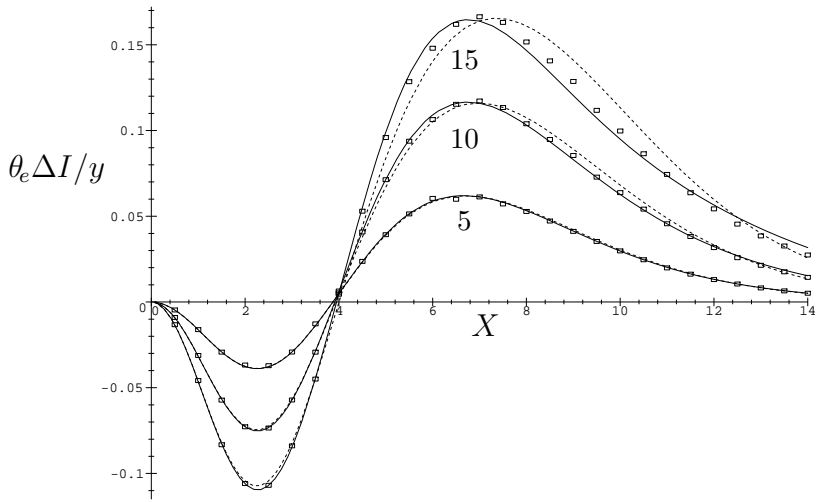,angle=-90,width=12cm}}}
\put(168,80){\makebox(0,0)[t]{\normalsize $X$}}
\put(24,135){\makebox(0,0)[r]{\normalsize $\thte\Delta I/y$}}
\put(160,119){\makebox(0,0)[t]{\small $5$}}
\put(160,150){\makebox(0,0)[t]{\small $10$}}
\put(160,180){\makebox(0,0)[t]{\small $15$}}
  
\end{picture}
\end{center}
\caption{\sl{The intensity change $\thte\Delta I/y$
(in units of $2(k_{B}T_{0})^{3}
/(hc)^{2}$) plotted against $X$ for three values
of $k_{B}T_{e}$ (in $\kev$).
The solid curves are calculated using the second-order
correction to the Kompaneets equation, while the dashed lines are
calculated from the first-order correction. The points are the results
of a Monte-Carlo evaluation of the Boltzmann collision integral.}}
\label{sz_fig_3}
\end{figure}
 
%%%%%%%%%%%%%%%%%%%%%%%%%%%%%%%%%%%%%%%%

In Figure~\ref{sz_fig_3} we compare the spectrum of $\Delta I$ calculated
with equation~\eqref{sz_eq_z4} to the spectrum with the
correction~\eqref{sz_eq_s_1} included, for $k_{B}T_{e} = 5$, $10$ and
$15\kev$ ($\thte \approx 0.01$,
$0.02$ and $0.03$ respectively). In each case, the
second-order relativistic effects are not significant in the Rayleigh-Jeans
part of the spectrum. This is to be expected from inspection
of equation~\eqref{sz_eq_s_2}, where the $\thte^{2}$ term is
clearly insignificant for the values of $\thte$ considered.
For $k_{B}T_{e}=5\kev$, the second-order effects are
insignificant over the entire spectrum. However, for $k_{B}T_{e} \simgt
10\kev$, the second-order effects make a significant contribution to
the relativistic correction to the Kompaneets based prediction outside the
Rayleigh-Jeans region. We have verified that the
third-order corrections are negligible over the entire spectrum for
$k_{B}T_{e} \simeq 10\kev$.
This is confirmed by a comparison of the curves in Figure~\ref{sz_fig_3}
with the points which are the results of a direct Monte-Carlo
evaluation of the Boltzmann collision integral with $n(\omega)$
given by the Planck distribution $n_{0}(x)$~\cite{gull97}. The
second-order effects should be included in the analysis of high frequency data
for hot clusters. The magnitude of the second-order
correction to the Sunyaev-Zel'dovich result for the rather mild
values of $\thte$ considered here, is symptomatic of the asymptotic nature of
the series expansion of $\partial n/\partial t$ in $\thte$.
However, for the majority of clusters
considered in Sunyaev-Zel'dovich analyses, the inclusion of the first
two relativistic corrections should be sufficient, particularly for
experiments working in the Rayleigh-Jeans region of the spectrum.

\subsection{The Crossover Frequency}

%%%%%%%%%%%%%%%%%%%%%%%%%%%%%%%%%%%
 
\begin{figure}[t!]
\begin{center}
\begin{picture}(340,210)
 
%\multiput(0,0)(10,0){35}{\line(0,1){210}}
%\put(0,0){\makebox(0,0)[t]{\tiny 0}}
%\put(50,0){\makebox(0,0)[t]{\tiny 50}}
%\put(100,0){\makebox(0,0)[t]{\tiny 100}}
%\put(0,0){\makebox(0,0)[r]{\tiny 0}}
%\put(0,50){\makebox(0,0)[r]{\tiny 50}}
%\put(0,100){\makebox(0,0)[r]{\tiny 100}}
%\multiput(0,0)(0,10){22}{\line(1,0){340}}
 
\put(-5,245){\hbox{\epsfig{figure=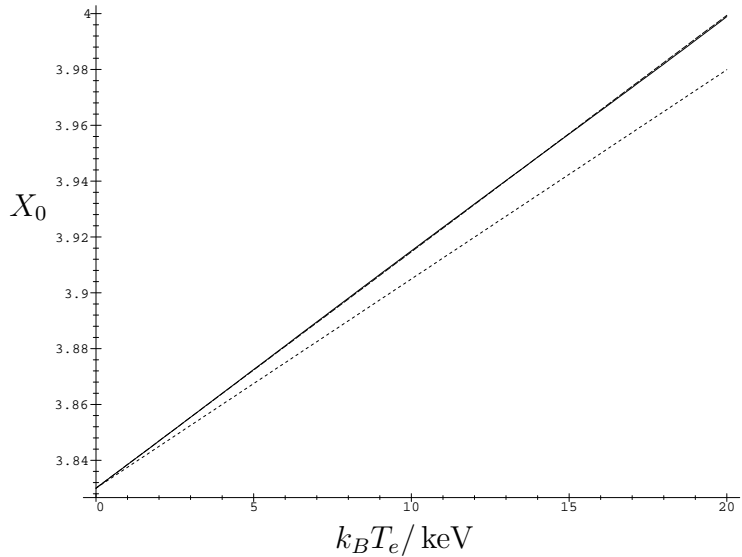,angle=-90,width=12cm}}}
\put(163,10){\makebox(0,0)[t]{\normalsize $k_{B}T_{e}/\kev$}}
\put(28,130){\makebox(0,0)[r]{\normalsize $X_{0}$}}
\end{picture}
\end{center}
\caption{\sl{The crossover frequency $X_{0}$ plotted against $k_{B}T_{e}$.
The solid line is calculated with the inclusion of third-order corrections
to the Kompaneets equation. The upper dotted line is a linear fit to the solid
line with $X_{0}=3.83(1+1.13\thte)$, while the lower dotted line is the
linear fit given in~\cite{reph95}: $X_{0} = 3.83(1+\thte)$.}}
\label{sz_fig_4}
\end{figure}
 
%%%%%%%%%%%%%%%%%%%%%%%%%%%%%%%%%%%%%%%%

The accurate determination of the crossover frequency $X_{0}$
(where the thermal component of the spectral distortion vanishes)
is essential for reliable
subtraction of the kinematic contribution to the Sunyaev-Zel'dovich
effect~\cite{reph95}.
In Figure~\ref{sz_fig_4} we plot the crossover frequency as a function of
$k_{B}T_{e}$, with the first three relativistic corrections included.
For $k_{B}T_{e} \simlt 20\kev$ we find that $X_{0}$ is well approximated by
the linear relation
\begin{equation}
X_{0} \simeq 3.83(1+1.13\thte).
\label{sz_eq_s_3}
\end{equation}
For comparison, Rephaeli~(1995),
found $X_{0}$ to be approximated by $X_{0}\simeq 3.83(1+\thte)$
in the interval $k_{B}T_{e} = 1 \mbox{--} 50\kev$, while Fabbri~(1981)
found $X_{0}\simeq 3.83(1+1.1\thte)$ for $k_{B}T_{e} \simlt 150\kev$.
It is clear that our calculation favours Fabbri's expression.
For $k_{B}T_{e} \simgt 20\kev$, $X_{0}$ calculated with the first three
relativistic corrections departs from the linear prediction~\eqref{sz_eq_s_3}.
However, we do not regard this as indicative of a breakdown of the linear
approximation, since it is clear from Figure~\ref{sz_fig_3}
that the inclusion of higher-order terms may have a significant effect on the
value of the crossover frequency.

\section{Conclusion}

We have shown how the Kompaneets equation may be generalised to include
relativistic effects in a self-consistent manner. The resulting equation
guarantees photon conservation at each order and is in agreement
with direct calculations
of the energy transfer rate between the plasma and a Planckian
distribution of photons.

We have applied this formalism to the calculation of the Sunyaev-Zel'dovich
effect in optically thin clusters. We presented simple analytic expressions
for the first two relativistic corrections to the usual Kompaneets based
expression for the spectral distortion, which are in excellent agreement
with the numerical calculations of Rephaeli~(1995) for electron
temperatures $\simlt 10\kev$.
This provides further evidence that the
low optical depth of clusters does not forbid the application of the
Boltzmann equation to the calculation of the Sunyaev-Zel'dovich
effect~\cite{fabbri81}.
The asymptotic nature of the expansion of $\partial n/\partial t$ in $\thte$
requires the inclusion of higher-order corrections
to calculate the effect in hotter clusters in the Wien region of the spectrum.
While the calculation of higher-order corrections is not problematic,
the bad convergence properties of the series means that ultimately one must
resort to a numerical calculation of the
collision integral~\cite{corman70} (or employ the multiple scattering
formalism~\cite{wright79}) to calculate the effect in very hot clusters.

Our calculations support the conclusions reached in~\cite{reph97}, that
including relativistic effects leads to a small decrease in the value of the
Hubble constant, inferred from combined X-ray and Sunyaev-Zel'dovich
information. For a cluster temperature of $\simeq 8\kev$, the reduction in
$H_{0}$ due to relativistic effects is $\simeq 5$ percent for measurements made
in the Rayleigh-Jeans region.

\acknowledgements

We would like to express our gratitude to Drs N. Itoh, Y. Kohyama and
S. Nozawa, whose preprint~\cite{itoh97} stimulated this research, and to
Drs S. Gull and A. Garrett for allowing us to quote the (unpublished) results
of a Monte-Carlo simulation.

\end{document}